\documentclass[a4paper]{jpconf}
\usepackage{amsmath}
\usepackage{bm}
\usepackage{wrapfig}
\usepackage{subfigure}
\usepackage[dvipdfmx]{graphicx}
\begin{document}
\title{Manipulation of the Dirac cones and the anomaly 
in the graphene related quantum Hall effect
}
\author{H Watanabe$^1$, Y Hatsugai$^2$ and H Aoki$^{1}$}
\address{$^1$ Department of Physics, University of Tokyo, Hongo, Tokyo 113-0033, Japan}
\address{$^2$ Institute of Physics, University of Tsukuba, Tsukuba, 305-8571, Japan}
\ead{watanabe@cms.s.t-tokyo.ac.jp}
\begin{abstract}
 The quantum Hall effect in graphene 
is regarded to be involving half-integer topological 
numbers associated with the massless Dirac particle, 
this is usually not apparent due to the doubling of 
the Dirac cones.  
Here we theoretically consider two classes of 
lattice models in which we manipulate the Dirac cones with either  (a) 
two Dirac points that have mutually different energies, or (b) multiple 
Dirac cones having different Fermi velocities.  
We have shown, 
with an explicit calculation of the topological (Chern) number 
for case (a) and with an adiabatic argument for case (b) 
that the results are consistent 
with the picture that a single Dirac fermion contributes
 the half-odd integer series
($\cdots$ -3/2, -1/2, 1/2, 3/2, $\cdots$) to the Hall conductivity 
when the Fermi energy traverses the Landau levels.
\end{abstract}

\section{Introduction}
In the graphene quantum Hall effect (QHE)\cite{Nov05,Kim05,NetoDirac},
 a most striking point is that the Hall conductivity 
$\sigma_{xy} = 2n+1 (n = 0, \pm1, ...)$ 
 in units of $-e^2/h$ 
with the spin degrees of freedom dropped here, 
becomes 
$\sigma_{xy} = n+1/2$ when we divide by two 
to have the contribution from each Dirac cone, 
since there are two of them at K and K' 
in graphene.   The appearance of doubled Dirac cones 
is consistent with the Nielsen-Ninomiya 
theorem\cite{Ninomiya}, which dictates 
that Dirac cones should appear in pairs
as far as the chiral symmetry is present.  
However, a natural question is: can we do better than 
just dividing by two to convince ourselves on the 
half-integer topological numbers.  

The quantization into half odd integers is 
of fundamental interest, but as far as 
lattice models are concerned, we always have a periodicity in the Brillouin 
zone, and the TKNN formula\cite{TKNN} dictates that 
integer topological numbers are imperative.  For instance, 
we can conceive lattice models that is outside  
the applicability of Nielsen-Ninomiya, but 
we still end up with integer Hall conductivity. 
Here we take another approach to explore this problem:  
We theoretically consider two classes of 
lattice models in which we manipulate the Dirac cones with either (a) 
the two Dirac points that have mutually different energies\cite{watanabe}, or (b) multiple 
Dirac cones with different Fermi velocities.  
 With explicit calculations of the topological (Chern) number, 
we shall show that the results for systematically manipulated 
Dirac cones are consistent 
with the picture that a single Dirac fermion contributes
 the half odd integer series
($\cdots$ -3/2, -1/2, 1/2, 3/2, $\cdots$) to the Hall conductivity 
when the Fermi energy traverses the Landau levels.

\begin{figure}[ht]
\begin{minipage}[ht]{0.39\hsize}
\subfigure[]{\includegraphics[width=80pt]{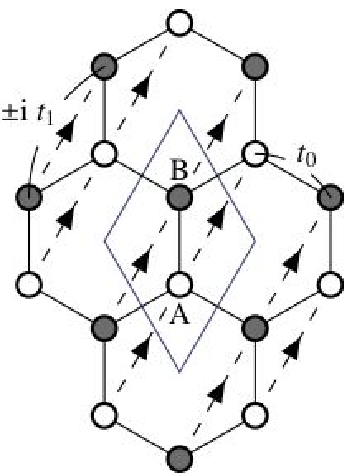}\label{fig:model}}
\end{minipage}
\begin{minipage}[ht]{0.59\hsize}
\subfigure[]{\includegraphics[width=145pt]{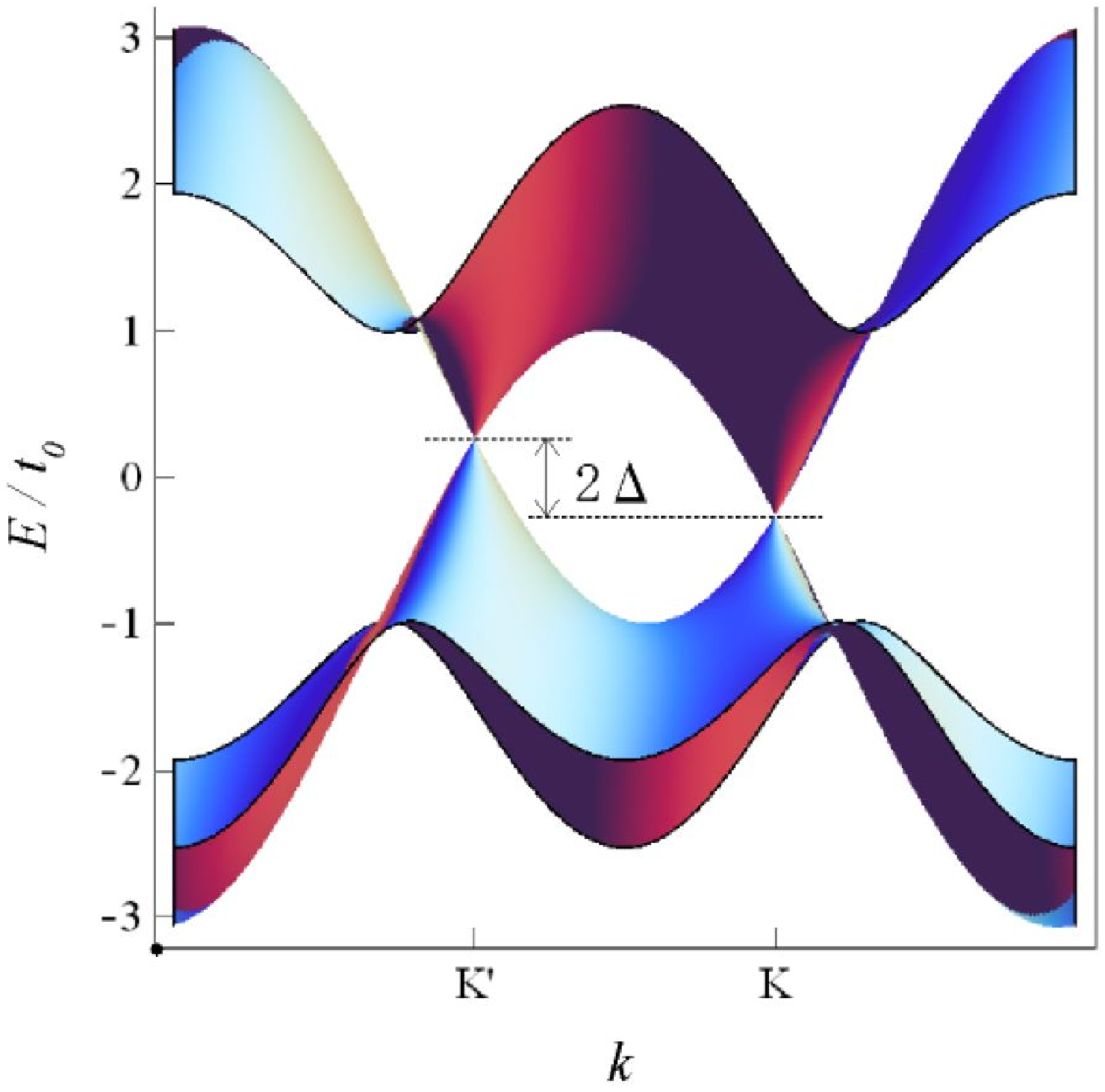}\label{fig:dirac}}
\end{minipage}
\caption{(a)The honeycomb lattice model considered here with 
nearest-neighbor hopping $t_0$ (solid lines) and 
second-neighbor hopping $i t_1$ along each 
dashed arrow, $-it_1$ in the opposite direction.  
A unit cell and A, B sublattices are indicated.  (b)
The dispersion relation in the present model, seen horizontally, 
with Dirac cones shifted in energy by $\Delta=\sqrt{3}t_1$.
}
\label{fig:dispersion}
\end{figure}

\section{Shifted Dirac cones}
A simplest way to 
realize  a model in which two Dirac cones are 
preserved but have 
shifted energies is to 
 add a term that is proportional to $\sigma_0$ (unit matrix) 
 with a $k$-dependent 
coefficient in the space spanned by Pauli matrices\cite{watanabe}.  
So we introduce a lattice model with a Hamiltonian, 
\begin{gather}
\label{eq:hamiltonian}
{\cal H} = \sum_{\bm{k}}\hat{c}^{\dagger}_{\bm{k}\alpha}[h^{\mathrm{gr}}_{\bm{k}}+2t_1(\sin k_1)\sigma_0]_{\alpha,\beta}\hat{c}_{\bm{k}\beta},\\
\notag
h^{\mathrm{gr}}_{\bm{k}}=
t_0\big[(1+\cos{k_1}+\cos{k_2})\sigma_1+(\sin{k_1}+\sin{k_2})\sigma_2\big],\end{gather}
where $\sigma_i$'s are Pauli matrices and $\alpha, \beta$ 
denote their components.  
Namely,  we have added, on top of the 
nearest-neighbor hopping $t_0$, an extra hopping 
involving $\sigma_0$ times a $\bm{k}$-dependent function.   
We then lift the 
degeneracy of the energies of K and K'  if the 
$\bm{k}$-dependent term has different values between them.  
A simplest choice is $\propto {\rm sin}k_1$.  
If we go back to the real space, the tight-binding model is as 
depicted in 
Fig.\ref{fig:model}, which has extra second-neighbor hoppings on top of the nearest-neighbor ones.   
The added hoppings has to be only between A-A and B-B 
for the Dirac cone to be preserved, and they have to 
be complex for the degeneracy between K and K', mutually time-reversal pairs, to be broken.  
The model with a complex hopping is fictitious, but 
we do accomplish shifted Dirac cones as depicted in Fig.\ref{fig:dirac}.  
In this model the chiral symmetry is broken, 
since $\sigma_0$-term in ${\cal H}$ invalidates 
$\{{\cal H}, \sigma_z\}=0$.   Nevertheless, the addition of 
$\sigma_0$ preserves the shape of Dirac cones, along with 
the species doubling.  
If we expand the Hamiltonian 
(\ref{eq:hamiltonian}) around $\bm{k}_0$ (K or K'), 
we have 
$
h_{\bm{k}}\simeq \chi\Delta\,\sigma_0
-\hbar v_F\big[(-1)^{\frac{1-\chi}{2}}\delta k_x\sigma_1+\delta k_y\sigma_2\big],$ 
where $\chi=+1(-1)$ correspond to K (K'), 
$\Delta=\sqrt{3}t_1$ is (half) the shift, $v_F=\sqrt{3}a\,t_0/2\hbar$ the Fermi velocity, 
and $\delta\bm{k} = \bm{k} - \bm{k}_0$.  

\section{Chern numbers} 
\begin{figure}[ht]
\begin{minipage}[ht]{0.49\hsize}
\includegraphics[width=220pt,clip]{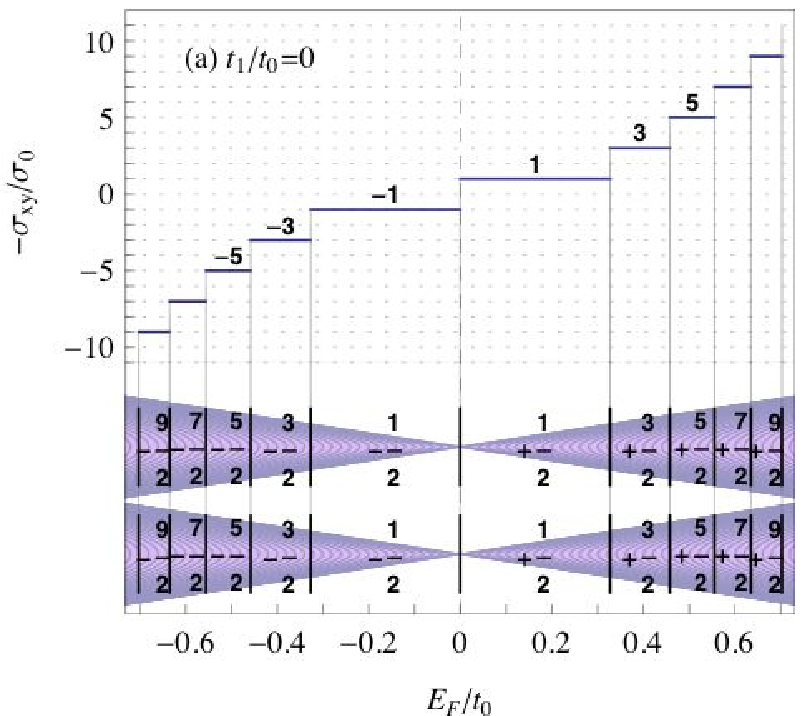}\label{fig:LLchern0}
\end{minipage}
\begin{minipage}[ht]{0.49\hsize}
\includegraphics[width=220pt,clip]{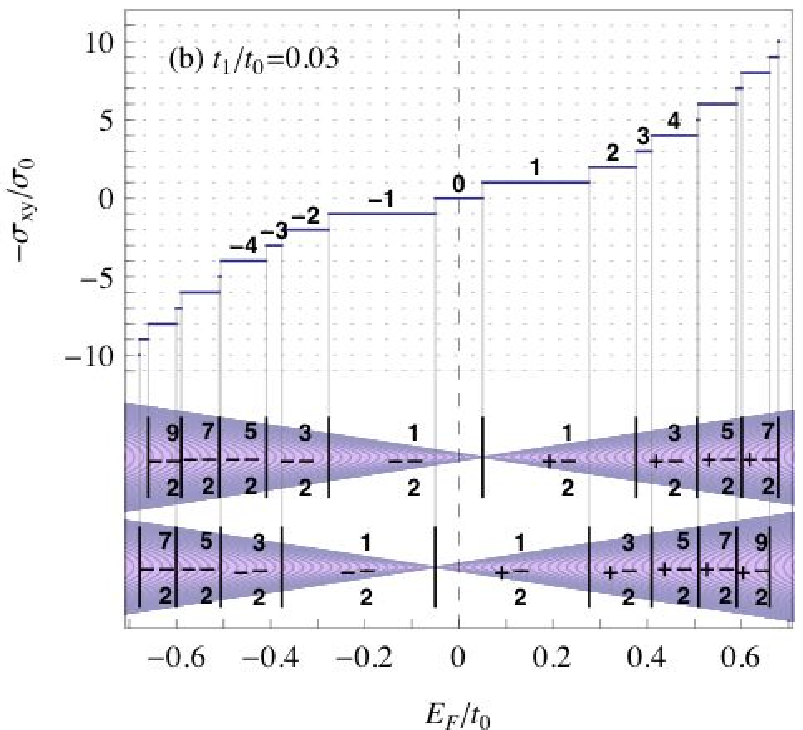}\label{fig:LLchern3}
\end{minipage}
\caption{Numerically calculated Chern numbers against $E_F$ 
(upper parts), along with the Landau levels for each of the 
two Dirac cones with the Chern numbers in each gap displayed (lower parts) for the ordinary graphene ($t_1=0$) 
(a) and for the present model with shifted Dirac cones ($t_1 \neq 0$) 
(b).  
The result is for the magnetic flux $\phi=1/100$.
\label{fig:LLchern}}
\end{figure}

The calculation of the Hall conductivity, which is a topological Chern 
number, has to be calculated carefully, since we have to sum up 
over the contribution from the ``Dirac sea".  
We can overcome the difficulty with a method employing non-commutative 
Berry's connection\cite{HatsugaiBerry} and its integration over the Brillouin 
zone to estimate the Chern number with a technique 
developed in the lattice gauge theory.\cite{Fukui}  
The result for the Chern number in the present model  
is shown in Fig.\ref{fig:LLchern}(b), while  Fig.\ref{fig:LLchern}(a) 
is for the  the ordinary graphene for comparison.  

Figure \ref{fig:LLchern}(b) shows that 
the result is exactly what we expect when 
we sum the 
two half-odd-integer series 
(..., -3/2, -1/2, 1/2, 3/2, …) with a shift in energy as 
$E_F$ traverses shifted sets of Landau levels 
 (as shown in the lower part in the figure).  
In other words, in a striking contrast to the 
ordinary graphene where each QHE step has a jump of 2 in 
the Chern number, the present model exhibits a jump of 1 
at each step\cite{watanabe}.  
The agreement is rather surprising (since there is no 
apriori reason why the superposition of effective field theory for the 
vicinities of K and K' and the lattice model should have the 
same Chern numbers).  
Thus, although we have still no half integers 
for the total Hall conductivity (since a sum of two half-odd 
integers is an integer), 
we have indirectly confirmed the half integers.  
Since no half integers should appear according to TKNN, 
this is indeed as best as we can confirm the half-integer 
property of each Dirac cone.

\section{Multiple Dirac cones as higher-pseudospin representations of SU(2)}

The effective theory for the Dirac cones in graphene is expressed 
in terms of Pauli matrices, which is a representation of 
SU(2) with pseudospin $1/2$.  Now we pose a question: can we extend 
this to examine whether the generalized model has 
half-integer contributions to the Chern number as well?  
For the two-band case (which may be regarded as 
having a pseudospin $1/2$) the each matrix is $2\times 2$, which corresponds to 
two (A, B) sublattices.  Here we show that a realization of 
higher pseudospin $S$ Dirac cones is possible, where each matrix is 
$(2S+1)\times (2S+1)$ corresponding to $(2S+1)$ sublattices.  
By a pseudospin $S$ Dirac cone we mean an effective Hamiltonian 
around the Dirac point,
\begin{equation}
\label{eq:highS}
{\cal H} = k_1\Sigma_1+k_2\Sigma_2+\mathcal{O}(k^2),
\end{equation}
where $\Sigma_i$'s are $(2S+1)$-dimensional representations 
of SU(2).   

One such model actually appears in the ``flat-band model", 
where one band is flat in a mult-band model.\cite{ando}  
Namely, in a Lieb's model for the flat band, $S=1$ SU(2) is 
realized for a three-band case, where two bands form a 
Dirac cone, while the other is flat.  We can generalize this 
to realize for general $S$ with $(2S+1)$ sites in a unit cell, 
although the models can again be unrealistic, but does serve as 
examining half-integer Chern numbers.   

\begin{figure}[!ht]
\begin{center}
\includegraphics[width=60pt,clip]{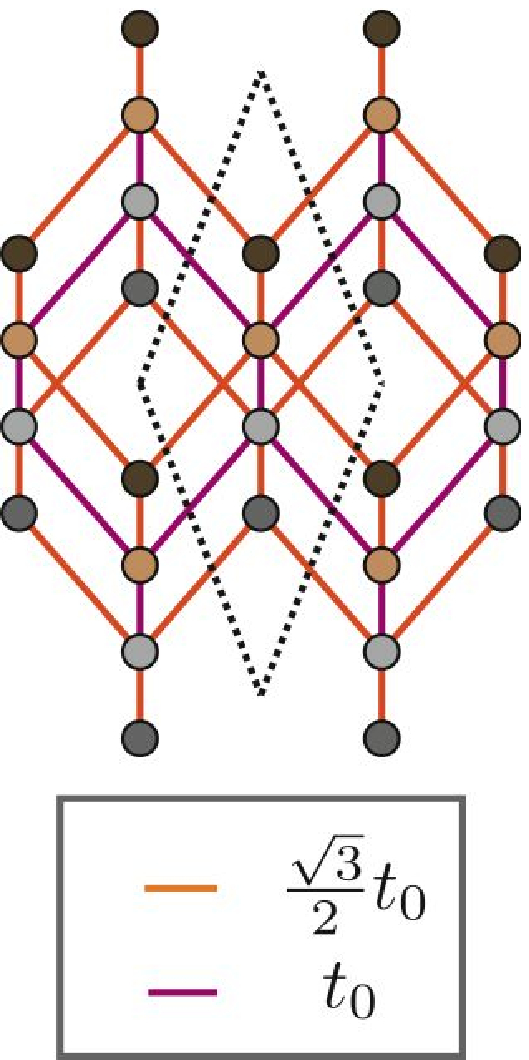}
\includegraphics[width=300pt,clip]{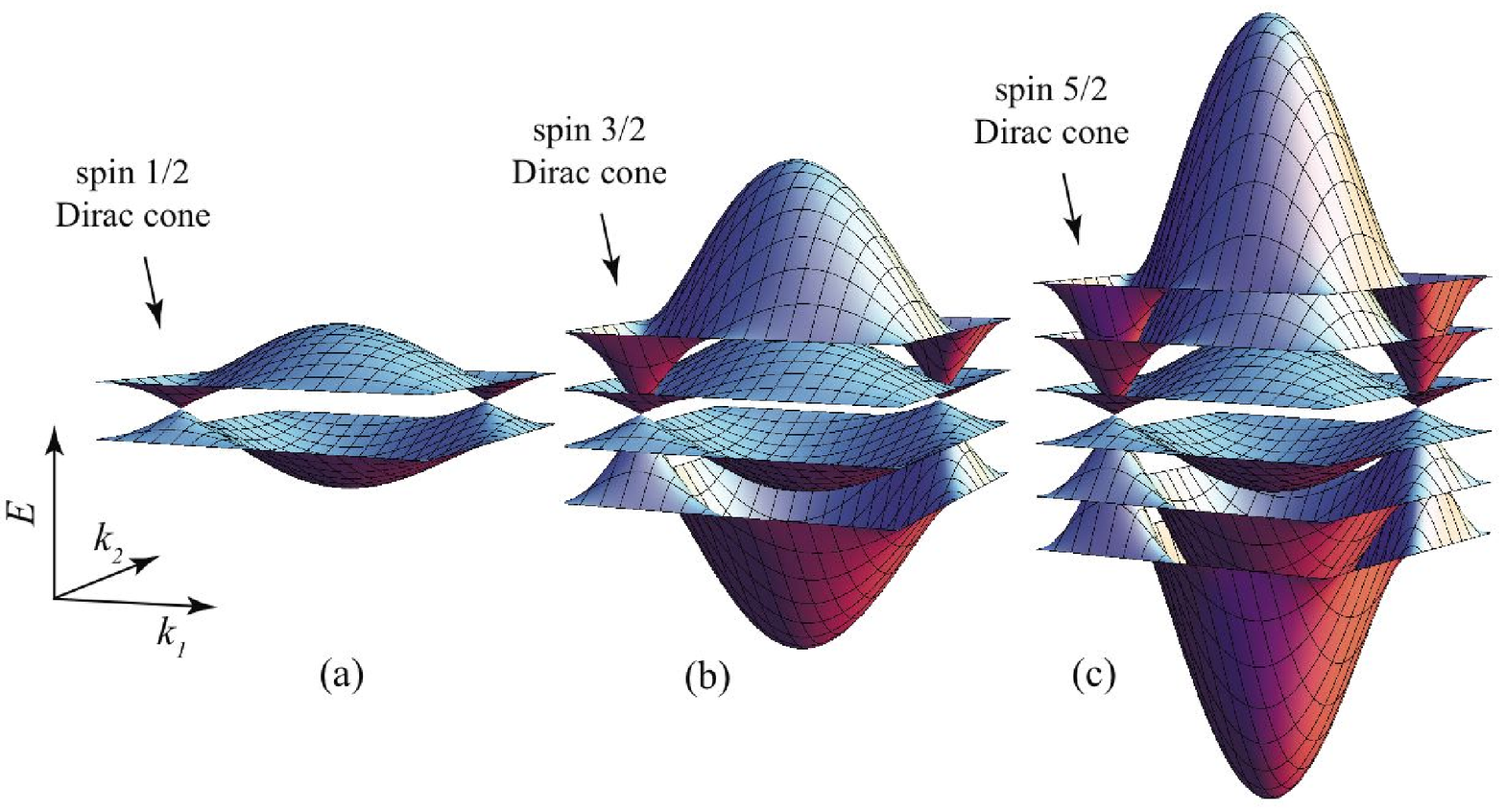}
\caption{Dispersions for the pseudospin $S$ SU(2) models that contain 
$(S+1/2)$ Dirac cones for $S=1/2$ (usual graphene) (a), 
$S=3/2$ (b), and $S=5/2$  (c). Inset depicts the lattice model for (b). }
\label{fig:dis}
\end{center}
\end{figure}

\begin{figure}[!ht]
\begin{minipage}[!ht]{0.45\hsize}
\includegraphics[width=150pt]{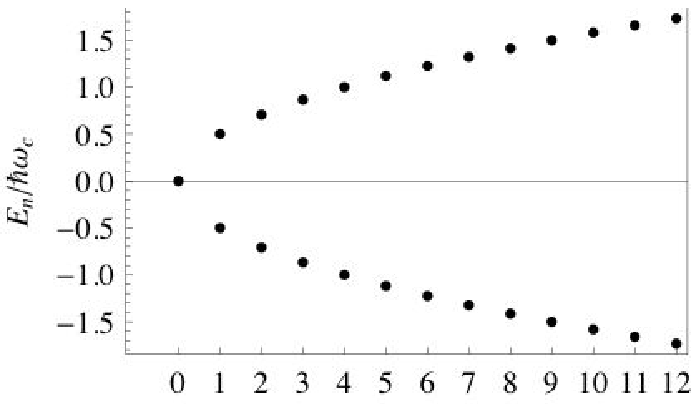}\label{fig:spin12}
\includegraphics[width=150pt]{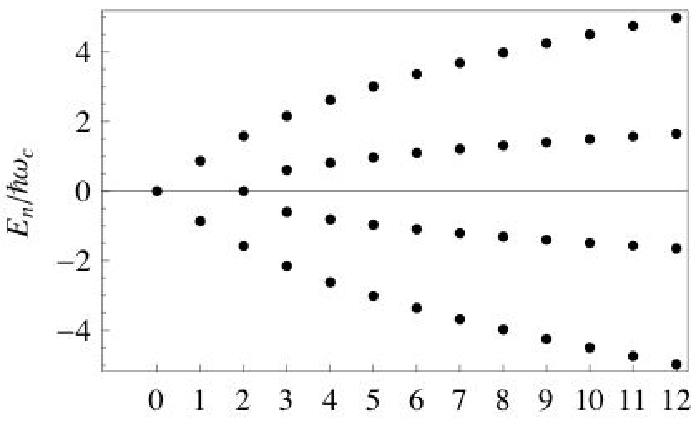}\label{fig:spin32}
\includegraphics[width=150pt]{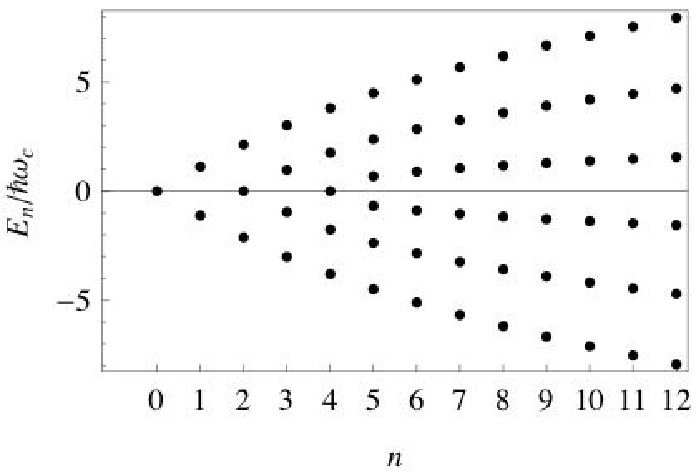}\label{fig:spin52}
\caption{Landau levels $E_n$ against $n$ for 
the pseudspin $S=1/2$ (ordinary graphene) (a), 
$S=3/2$ (b), and 
$S= 5/2$ (c).  
}
\label{fig:ll}
\end{minipage}
\hspace{30pt}
\begin{minipage}[!ht]{0.45\hsize}
\includegraphics[width=150pt,clip]{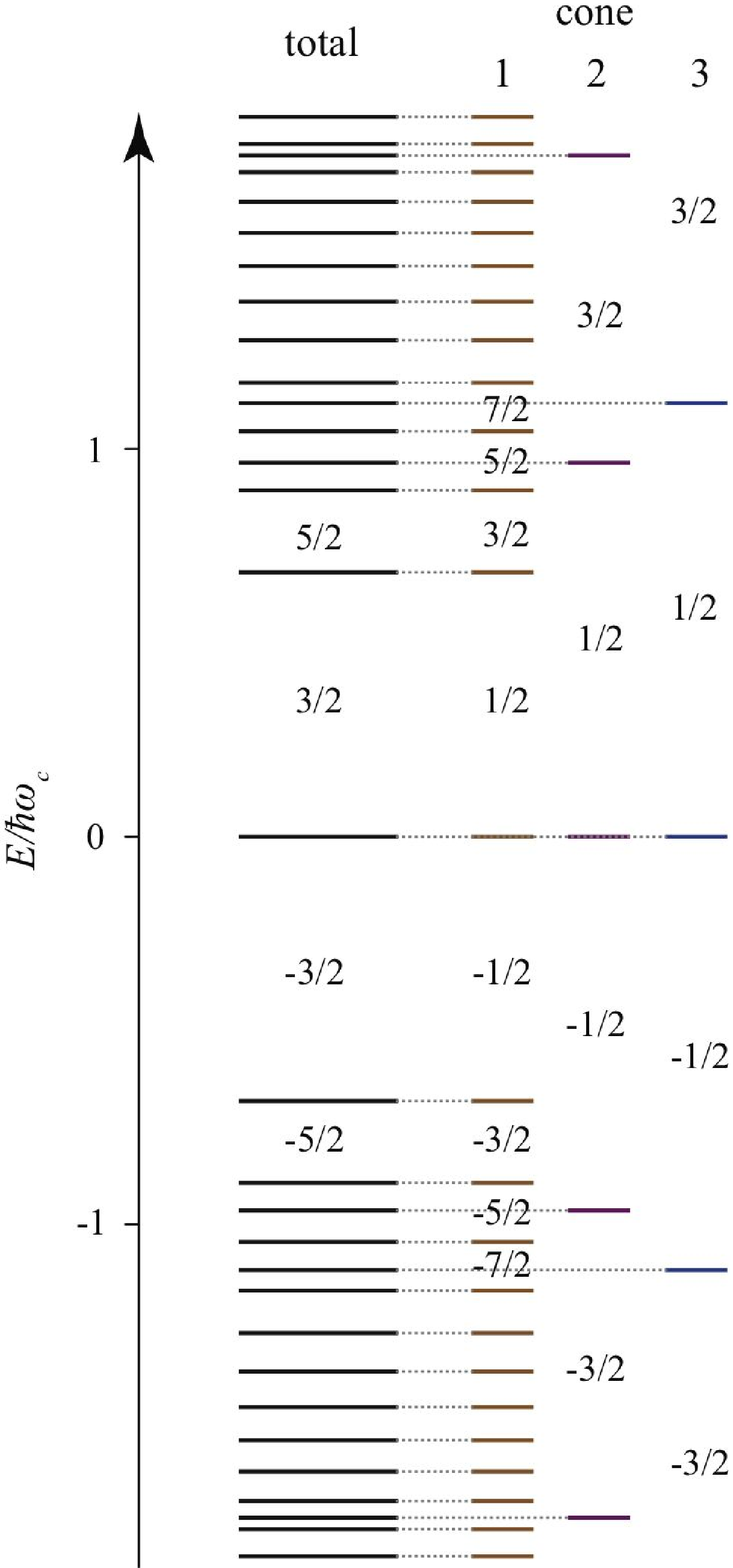}
\caption{The Hall conductivity (left column) 
and its decomposition into the contributions from 
three Dirac cones (right).}
\label{fig:hall}
\end{minipage}
\end{figure}

Although general values of $S$ are permissible, here we focus on half-integer $S=1/2, 3/2, 5/2, \cdots$ with $2S+1$ even 
(while the case of an integer $S$ contains a flat band in the band structure). 
We can construct the lattice models that realize these pseudo-spins by inserting extra atoms on edges of each hexagon with real hoppings between them. 
In this case $(2S+1)/2$ cones appear around each of K and K' points as displayed in Fig. \ref{fig:dis}. 
Let us first look at the Landau level structure ($E_n$ against $n$) in Fig. \ref{fig:ll}. 
This can be obtained by quantizing eqn. (\ref{eq:highS}) with $k$ replaced with $k +eA$. 
The Landau levels have $(2S+1)$ sequences that correspond to electron and hole Landau levels originating from the $(S+1/2)$ Dirac cones with different Fermi velocities, 
and they asymptotically approach $\hbar\omega_c\sqrt{n} S_z\,,\,\,(S_z=-S, -S+1, \cdots, S)$ for large $n$. 
The question is the Hall conductivity contributed by the multiple Dirac cones around each of K and K'. 
Figure \ref{fig:hall} displays the result for $S=5/2$, which is obtained from an effective field theory as follows. 
We first add a mass term $m\Sigma_3$ to calculate the Chern number in zero magnetic field, following the idea of  Haldane\cite{Haldane}, and then apply a magnetic field to have Landau levels and let $m\rightarrow 0$. 
Topological numbers should be invariant in such an adiabatic process. 
In Fig. \ref{fig:hall} for the contribute to the total  Chern numbers from the  three sets of Landau levels 
associated with the three Dirac cones with different Fermi velocities, we do have half-odd integer series, $\cdots, -5/2, -3/2, 3/2, 5/2, \cdots$ 
as another generalization of the half-integer contribution from each Dirac cone.  
For the original lattice model we have two Diract points, so we can apply the shift between the two Dirac points to resolve them as discussed in the first half of this paper, which is a future problem. 

Recently it is recognized\cite{HasanKane} that massless Dirac fermions can be realized as surface states of the three-dimensional topological 
(quantum spin Hall) systems as one manifestation of the bulk-edge correspondence\cite{HatsugaiEdge93,HFA}. 
There the doubling partner exists at the other side of the system, so that the decomposition of the topological numbers into contributions from each Dirac 
cone may become a realistic as well as intriguing question. The work was supported in part by grants-in-aid for scientific research 
No. 20340098 (YH and HA) from JSPS and No. 22014002(YH) on priority areas from MEXT.

\end{document}